\documentclass[twocolumn,english,two column,prl, Show PACS]{revtex4}
\usepackage[latin9]{inputenc}
\usepackage{float}
\usepackage{dvipost}
\usepackage{amsmath}
\usepackage{amssymb}
\usepackage{graphicx}

\makeatletter

\dvipostlayout
\dvipost{osstart color push Red}
\dvipost{osend color pop}
\dvipost{cbstart color push Blue}
\dvipost{cbend color pop}

\@ifundefined{textcolor}{}
{%
 \definecolor{BLACK}{gray}{0}
 \definecolor{WHITE}{gray}{1}
 \definecolor{RED}{rgb}{1,0,0}
 \definecolor{GREEN}{rgb}{0,1,0}
 \definecolor{BLUE}{rgb}{0,0,1}
 \definecolor{CYAN}{cmyk}{1,0,0,0}
 \definecolor{MAGENTA}{cmyk}{0,1,0,0}
 \definecolor{YELLOW}{cmyk}{0,0,1,0}
}

\usepackage{epsfig}\usepackage{bm}\usepackage{amsfonts}

\makeatother

\usepackage{babel}
\begin{document}

\title{Filtration and Extraction of Quantum States from Classical Inputs}

\author{Chang-Ling Zou$^{1,2,3}$ }

\author{Liang Jiang$^{2}$ }

\email{liang.jiang@yale.edu}

\author{Xu-Bo Zou$^{1,3}$ }

\email{xbz@ustc.edu.cn}

\author{Guang-Can Guo$^{1,3}$}

\address{$^{1}$Key Lab of Quantum Information, University of Science and
Technology of China, Hefei 230026 }

\address{$^{2}$Department of Applied Physics, Yale University, New Haven,
CT 06511, USA }

\address{$^{3}$Synergetic Innovation Center of Quantum Information \& Quantum
Physics, University of Science and Technology of China, Hefei, Anhui
230026, China}

\date{\today }
\begin{abstract}
We propose using nonlinear Mach-Zehnder interferometer (NMZI) to efficiently
prepare photonic quantum states from a classical input. We first analytically
investigate the simple NMZI that can filtrate single photon state
from weak coherent state by preferrentially blocking two-photon component.
As a generalization, we show that the cascaded NMZI can deterministically
extract arbitrary quantum state from a strong coherent state. Finally,
we numerically demonstrate that the cascaded NMZI can be very efficient
in both the input power and the level of cascade. The protocol of
quantum state preparation with NMZI can be extended to various systems
of bosonic modes.
\end{abstract}


\maketitle
\emph{Introduction.- }Integrated photonics can achieve unprecedented
interferometric stability \cite{politi2008silica,OBrien2009} and
build large scale interferometers \cite{Carolan2015,Harris2015}.
However, reliable quantum state preparation for integrated photonics
remains an important challenge, because interferometers and coherent
input states are insufficient for quantum states preparation. We may
use either post-selection or nonlinear interaction to overcome this
challenge. The approach of post-selection only requires linear optical
elements and photon detectors, but the preparation of quantum state
is probabilistic and conditioned on the outcome of the projective
measurement \cite{Kok2007,Chrzanowski2014,knill2001scheme}. The approach
of nonlinear interaction assisted by an ancillary two-level system
(TLS) can deterministically prepare arbitrary quantum state of the
photonic mode \cite{Law1996,Ben-Kish2003,Hofheinz2009}, but it requires
strong coupling between the optical mode with the single TLS, which
is experimentally challenging for integrated photonics. Alternatively,
we may consider using the nonlinear optical waveguide combined with
ultra-stable interferometers to achieve reliable quantum state preparation,
without requiring TLS \cite{Law1996,Ben-Kish2003,Hofheinz2009}, post-selection
\cite{Zou2002,Hofmann2002,Sanaka2006}, nor feedback/feedforward control
\cite{Sayrin2011}.

\begin{figure}[ptb]
\centerline{ \includegraphics[width=0.45\textwidth]{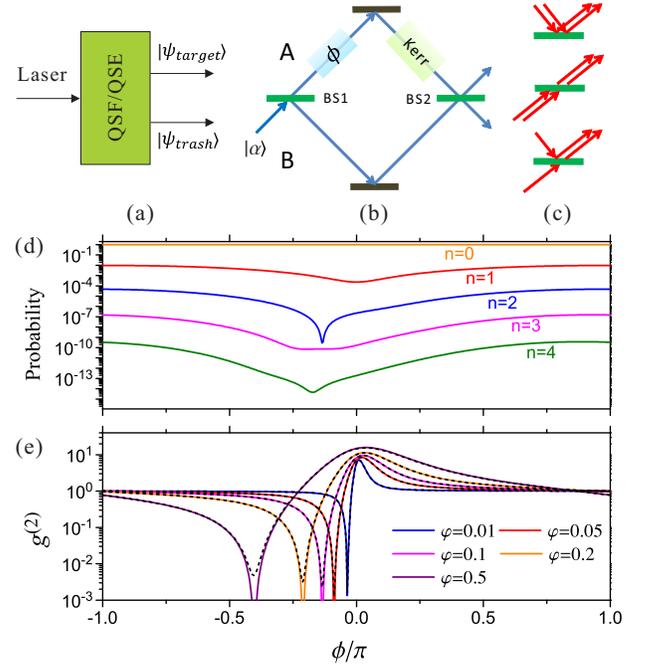}}\protect\caption{(color online) (a) Schematic illustration the arbitrary quantum state
filtration and extraction from coherent state input. (b) The configuration
for QSF of single photon from coherent state input $\left|\alpha\right\rangle $,
using the simple NMZI (consisting of Mach--Zehnder interferometer,
and Kerr medium and phase shifter). (c) Three processes for two photon
output of Path A for weak coherent input. (d) The probabilities of
$n$ photons output of Path A against the phase difference between
two arms $\phi$, with $\varphi=0.1$ and $\alpha=0.1$. (e) The second-order
correlation function ($g^{(2)}$) of light output of Path A against
$\phi$ for various $\varphi$ with $\alpha=0.1$. }

\label{singlephoton}
\end{figure}

In this Letter, we propose to use interferometry combined with Kerr
nonlinearity to filtrate single photons or extract any desired quantum
states from coherent state input, as illustrated in Fig.\ \ref{singlephoton}(a).
We first present the idea of quantum state filtration (QSF) of single
photons, which keeps the desired single photon component by \textit{blocking}
the undesired component to a different port. We then generalize the
idea to quantum state extraction (QSE), which not only keeps the desired
component, but also extracts the desired component from the undesired
component before blocking/redirecting the residual photons.

\emph{Single photon filtration.-} We first consider the simple task
of QSE of single photons. As shown in Fig.\ \ref{singlephoton}(b),
we use a nonlinear Mach--Zehnder interferometer (NMZI), with a Kerr
nonlinear medium in one of the arms. Since Kerr nonlinearity can induce
photon number dependent phase shift, we can design the NMZI to induce
destructive interference at the output port when there are two photons.
More specifically, with a vacuum input at Path A (upper path) and
a coherent state input at Path B (lower path), the input state to
the filtration is
\begin{equation}
\left\vert \psi\right\rangle _{in}=\left\vert vac\right\rangle _{A}\otimes\left\vert \alpha\right\rangle _{B},
\end{equation}
where $\left\vert \alpha\right\rangle =e^{-\left\vert \alpha\right\vert ^{2}/2}\sum_{n=0}^{\infty}\frac{\alpha^{n}}{n!}(b^{\dagger})^{n}\left\vert vac\right\rangle $,
and $a(a^{\dagger})$ and $b(b^{\dagger})$ are annihilation (creation)
operators for Paths A and B, respectively. Each beam splitter (BS)
induces a unitary evolution,
\begin{equation}
U_{BS}(\theta_{1,2})=e^{i\theta_{1,2}(a^{\dagger}b+ab^{\dagger})},
\end{equation}
with $\theta_{1}$ and $\theta_{2}$ for BS1 and BS2, respectively.
The evolution in the nonlinear Kerr medium in Path A is
\begin{equation}
U_{K}(\phi,\varphi)=e^{i\phi a^{\dagger}a+i\varphi a^{\dagger}a^{\dagger}aa},
\end{equation}
where $\varphi$ is the Kerr coefficient and $\phi$ is linear phase
shift (relative to Path B). The final output state of the single photon
filtration is
\begin{align}
\left\vert \psi\right\rangle _{out}= & U_{BS}(\theta_{2})U_{K}(\phi,\varphi)U_{BS}(\theta_{1})\left\vert \psi\right\rangle _{in}\nonumber \\
= & \sum_{p=0}^{\infty}\sum_{q=0}^{\infty}\mu_{p,q}(a^{\dagger})^{p}(b^{\dagger})^{q}\left\vert vac\right\rangle ,
\end{align}
with $\mu_{p,q}=\sum_{l=0}^{p}\sum_{n=l}^{l+q}\lambda_{n,p+q-n}\binom{n}{l}\binom{p+q-n}{p-l}\times(-1)^{n-l}(\mathrm{sin}\theta_{2})^{p+n-2l}(\mathrm{cos}\theta_{2})^{q-n+2l}$.
The probability of $p$ photons at the output of Path A is
\[
P_{p}=\left\langle p\right\vert \mathrm{Tr}_{B}\{\left\vert \psi_{2}\right\rangle \left\langle \psi_{2}\right\vert \}\left\vert p\right\rangle =\sum_{q=0}^{\infty}p!q!\left\vert \mu_{p,q}\right\vert ^{2}.
\]
The second-order correlation function \cite{Agarwal2012} is
\begin{equation}
g^{(2)}=\frac{\left\langle a^{\dagger}a^{\dagger}aa\right\rangle }{\left\langle a^{\dagger}a\right\rangle ^{2}}=\frac{\sum_{p=2}^{\infty}p(p-1)\times P_{p}}{(\sum_{p=1}^{\infty}p\times P_{p})^{2}},\label{eq-complete}
\end{equation}
which characterizes the generated single photon state.

For a weak coherent input $\left|\alpha\right|^{2}\ll1$, we have
$P_{2}\ll P_{1}$ and can safely neglect the probability of multiple
photons ($P_{n\ge3}$). By considering the leading contribution, we
have
\begin{equation}
g^{(2)}\approx\frac{2P_{2}}{P_{1}{}^{2}}\approx\left\vert \frac{2\mu_{2,0}}{\mu_{1,0}^{2}}\right\vert ^{2}=\left\vert 1-\frac{1-e^{i2\varphi}}{(1-\eta e^{-i\phi})^{2}}\right\vert ^{2},\label{eq-approx}
\end{equation}
where $\mu_{1,0}=\alpha\mathrm{cos}\theta_{1}\mathrm{cos}\theta_{2}(e^{i\phi}-\eta)$
and $\mu_{2,0}=\frac{1}{2}(\alpha\mathrm{cos}\theta_{1}\mathrm{cos}\theta_{2})^{2}[-2\eta e^{i\phi}+e^{i2(\phi+\varphi)}+\eta^{2}]$
for the simple NMZI with $\eta=\mathrm{tan}\theta_{1}\mathrm{tan}\theta_{2}$.
The three terms in $\mu_{2,0}$ correspond to three different processes
with two photons at the output of Path A, as shown in Fig.\ \ref{singlephoton}(c).
The interference of these three processes can be controlled by the
linear phase shift ($\phi$) and nonlinear coefficient ($\varphi$).
The optimal condition for $\mu_{2,0}=0$ is
\begin{equation}
\eta e^{-i\phi}=1\pm\sqrt{1-e^{i2\varphi}},\label{eq-optimal}
\end{equation}
which can always be fulfilled as long as $\varphi\neq0$, so that
the leading contribution to $g^{(2)}$ can be eliminated.

Fig.\ \ref{singlephoton}(d) shows the probability of $n$ photons
at the output of Path A ($P_{n}$) depending on the linear phase shift
$\phi$, with parameters $\varphi=0.1$, $\eta=|1-\sqrt{1-e^{i2\varphi}}|$
and $\alpha=0.1$. We find that the $P_{2}$ is greatly suppressed
for $\phi\approx0.13\pi$, while the dominant single photon emission
$P_{1}\gg P_{2,3,4}$ is not significantly affected. In Fig.\ \ref{singlephoton}(e),
the relation between $g^{(2)}$ and $\phi$ are plotted for different
values of nonlinear coefficient $\varphi$, with $\alpha=0.1$ and
$\eta$ given by optimal condition from Eq.\ (\ref{eq-optimal}).
We find good agreement between the approximated analytical solution
from Eqs.\ (\ref{eq-approx})\&(\ref{eq-optimal}) (solid lines)
and the exact numerical solution from Eq.\ (\ref{eq-complete}) (dashed
lines). With increasing nonlinear coefficient $\varphi$, the deviation
from $g^{(2)}=1$ becomes more significant, due to the Fano interference
of the three processes (Fig.\ \ref{singlephoton}(c)) contributing
to $\mu_{2,0}$.  These Fano-like curves show sub-Poisson statistic
with $g^{(2)}\approx0$ for $\phi$ close to the optimal condition
(Eq.~(\ref{eq-optimal})), where the two photon output can be totally
forbidden due to destructive interference. Meanwhile, we can also
find the constructive interference of the two-photon output, which
gives rise to super-Poisson statistic ($g^{(2)}(0)\gg1$) output.
Comparing the curves with different nonlinear effect coefficients,
the single photon filtration is more sensitive to phase $\phi$ for
smaller $\varphi$, indicating the crucial role of nonlinearity.

For QSF of single photon, the fidelity is $F=P_{1}=(\mathrm{cos}\theta_{1}\mathrm{cos}\theta_{2})^{2}|\alpha^{2}|\left|1-e^{i2\varphi}\right|$.
The optimal condition requires $\eta=\mathrm{tan}\theta_{1}\mathrm{tan}\theta_{2}\approx1$,
we have $\left|\mathrm{cos}\theta_{1}\mathrm{cos}\theta_{2}\right|<\frac{1}{2}$
and $P_{1}<\varphi|\alpha^{2}|/2$, which implies that the fidelity
depends on both the the Kerr nonlinearity coefficient and the intensity
of the coherent state input. QSF with simple NMZI cannot suppress
the components with $n>2$ photons (see Fig. \ref{singlephoton}(d)
and also Ref. %
\footnote{See the Supplemental Material for additional details about the limitation
of two-unit NMZI QSF, performance of QSF/QSE for various Kerr nonlinearities
and imperfection.%
}), and it only works for weak coherent state $\left|\alpha^{2}\right|\ll1$,
which significantly limits the fidelity. Moreover, the fidelity of
QSF is fundamentally limited by the overlap between the input state
and the target state, $P_{succ}<\left|\left\langle \psi_{out}|\psi_{in}\right\rangle \right|^{2}$,
because it blocks all undesired components. To go beyond this limit,
we need to generalize QSF to QSE, which not only keeps the desired
component, but also extracts the desired component from the undesired
ones.

\begin{figure}[ptb]
\centerline{ \includegraphics[width=0.45\textwidth]{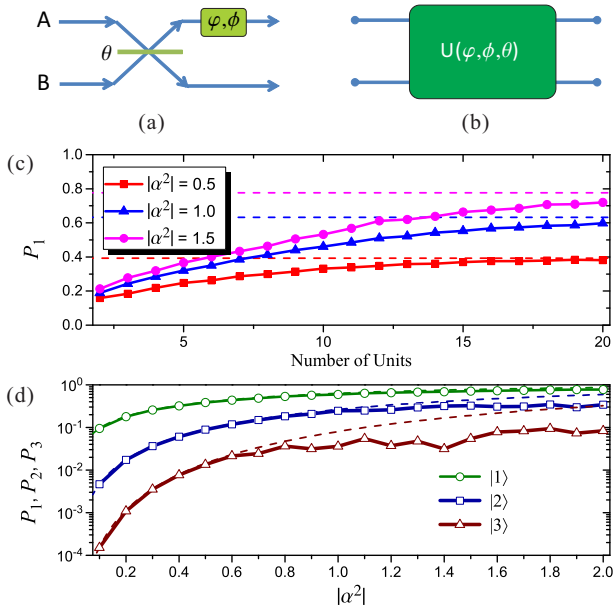}}

\protect\caption{(color online) Cascaded NMZI. (a) The basic element consists of a
BS ($\theta$) followed by a linear phase shifter ($\phi$) and a
Kerr medium ($\varphi$) in the upper path. (b) Schematic representation
of the element for cascaded NMZI. (c) Fidelity of the single photon
extraction increases with the number of cascade elements, $N$. The
parameters are optimized numerically under the constrain $P_{n\geq2}<0.01$
with $\varphi=0.1$. (d) Fidelity of Fock state extraction ($n=1,2,3$)
increase with $\left|\alpha\right|^{2}$ for cascaded NMZI with $N=40$.
The results are obtained by optimize the parameters of each unit under
the constrain that $1-P_{0}-P_{n}\leq0.01$.}

\label{Cascade}
\end{figure}

\emph{Cascaded filtration.-} To implement QSE, we consider the cascaded
NMZI, with a series of NMZIs connected sequentially. As shown in Fig.\ \ref{Cascade}(a),
the basic element consists of a BS ($\theta$) followed by a linear
phase shifter ($\phi$) and a Kerr medium ($\varphi$) in the upper
path. The basic element can be represented by a standard two-port
unitary {[}Fig.\ \ref{Cascade}(b){]}
\begin{equation}
U(\phi,\varphi,\theta)=U_{\mathrm{K}}(\phi,\varphi)U_{BS}.
\end{equation}
The cascaded MNZI with $N$ elements can be characterized by
\begin{equation}
U_{N}=\Pi_{l=1}^{N}U(\phi_{l},\varphi_{l},\theta_{l}).
\end{equation}
For example, the simple NMZI (Fig\ \ref{singlephoton}(b)) consists
of $N=2$ basic elements, with $\phi_{2}=\varphi_{2}=0$.

The cascaded NMZI can not only keep the desired single-photon component,
but also extract the (desired) single-photon state from (undesired)
multi-photon states, as long as there are enough photons in the undesired
component. We numerically optimize the fidelity by tuning the parameters
of the N elements. As illustrated in Fig.\ \ref{Cascade}(c), the
optimized fidelity of single photon extraction $F=P_{1}$ increases
with $N$ monotonically, with asymptotic value $F\rightarrow1-\left|\left\langle 0|\alpha\right\rangle \right|^{2}$
(dashed lines), because our passive device cannot extract single photon
from the vacuum component. Furthermore, the cascaded NMZI can extract
Fock state $\left\vert n\right\rangle $ with $n=1,2,3,\cdots$. The
asymptotic fidelity of $n$-photon extraction is $F\rightarrow1-\sum_{m=0}^{n-1}\left|\left\langle m|\alpha\right\rangle \right|^{2}$,
which can be achieved for $|\alpha|^{2}\leq1.5/n$ with cascaded NMZI
of $N=40$ elements, as shown in Fig.\ \ref{Cascade}(d).

\emph{Arbitrary state Extraction.-}Remarkably, the cascaded NMZI
can extract \textit{arbitrary} superposition of Fock states with a
large coherent state input ($|\alpha|\gg1$) with almost perfect fidelity.
For $\theta_{l}\ll1$ with $l=1,\cdots,N$, almost all input photons
will be guided in Path B, which effectively remains as a coherent
state (with small deviation of $O\left(\theta\right)$) for all intermediate
stages. The effect of each beam splitter to the upper path can be
regarded as an effective displacement operation to Path A, as $D(\epsilon_{l})=e^{\epsilon_{l}a^{\dagger}-\epsilon_{l}^{*}a}$
with $\epsilon_{l}=\alpha\theta_{l}$ and a small deviation of $O\left(\epsilon_{l}^{2}/\alpha^{2}\right)$
\cite{Paris1996}. In addition, the linear phase shift and Kerr nonlinearity
can achieve the unitary evolution $U_{K}\left(\phi_{l},\varphi\right)=e^{i\phi_{l}a^{\dag}a+i\varphi a^{\dag}a^{\dag}aa}$.
Hence, the cascaded NMZI of $N$ elements can induce the unitary evolution
$U_{K}\left(\phi_{N},\varphi\right)U\left(\epsilon_{N}\right)\cdots U_{K}\left(\phi_{2},\varphi\right)U\left(\epsilon_{2}\right)U_{K}\left(\phi_{1},\varphi\right)U\left(\epsilon_{1}\right)$,
which in principle can accomplish any desired unitary transformation
for sufficiently large $N$ and carefully chosen $\left\{ \phi_{l},\epsilon_{l}\right\} _{l=1,\cdots,N}$
\cite{Deutsch1995,Lloyd1995,Braunstein2005}. Despite the large overhead
in $N$, this provides a generic approach using cascaded NMZI to extract
arbitrary superposition of Fock states from a large coherent state
with almost perfect fidelity.

\begin{figure}[ptb]
\centerline{ \includegraphics[width=0.45\textwidth]{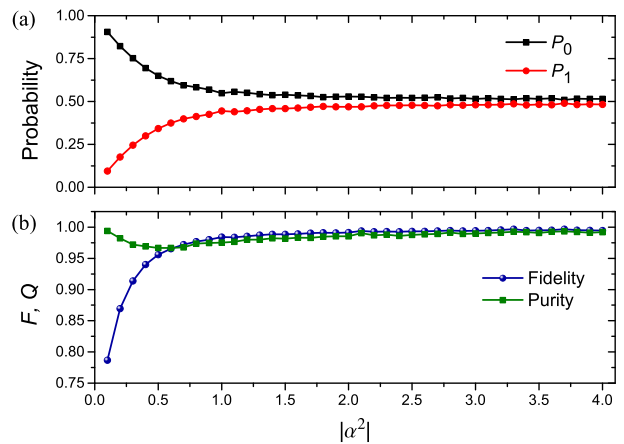}}\protect\caption{QSE for $\left\vert \psi_{target}\right\rangle =(\left\vert 0\right\rangle +\left\vert 1\right\rangle )/\sqrt{2}$.
(a) The probability of $\left\vert 0\right\rangle $ and $\left\vert 1\right\rangle $,
(b) the Fidelity $(F)$ and Purity ($Q$) of the output. The results
are obtained by optimize the parameters of a chain of $N=20$ units
under the constrain that $1-P_{0}-P_{1}\leq0.01$.}

\label{QSE}
\end{figure}

In practice, it is favorable to design the cascaded NMZI with a small
number of elements. To illustrate the feasibility, we consider the
target state $\left\vert \psi_{target}\right\rangle =(\left\vert 0\right\rangle +\left\vert 1\right\rangle )/\sqrt{2}$
using $N=20$ cascaded elements optimize the fidelity by tuning parameters
of $\left\{ \varphi_{l},\phi_{l},\theta_{l}\right\} _{l=1,\cdots,N}$.
As illustrated in Fig.~\ref{QSE}, we can improve the fidelity $F$
and purity $Q=\mathrm{Tr}(\rho_{A}^{2})$ of the extracted state by
increasing $\left|\alpha\right|^{2}$. Both $F$ and $Q$ are greater
than $97.5\%$ when $|\alpha|^{2}\ge1.5$. It's intriguing that a
high fidelity QSE of superposition of Fock states can be achieved
using a reasonable size coherent state and a finite-stage cascaded
NMZI.

\emph{Discussion.- }The photonic integrated circuits (PICs) provides
a promising platform for realizing cascaded NMZI, where arrays of
beam splitters and phase shifters can be integrated on a chip \cite{Carolan2015,Harris2015}.
QSE can provide arbitrary input photon states for quantum information
processing \cite{knill2001scheme,politi2008silica}. For experiment
realization, the most challenging part is the Kerr nonlinear at single
photon level ($\varphi=0.1$ in this paper). Compared with schemes
using single TLS at near-resonance condition, the Kerr nonlinearity
can be realized by collective effect of ensembles and is more feasible
for experiments. Recently, the interface between atomic ensemble and
photonic waveguide for single photon nonlinearity, such as nanofiber
\cite{Hendrickson2010,Vetsch2010}, hollow-core photonic crystal fiber
\cite{Bajcsy2009,Venkataraman2011} and integrated waveguide \cite{Stern2013},
have been deonstrated \cite{Chang2014,Tiecke2014,Goban2014}. Novel
mechanisms including giant Kerr effect \cite{Schmidt1996} and strong
interaction between Rydberg atoms \cite{Peyronel2012} have also been
proposed and demonstrated. As the integrated photonic chips starting
the new era of fJ-level (1000 photons) nonlinear effect \cite{Kwon2013,Notomi2014},
new materials, such as graphene \cite{Gullans2013} and topological
insulator \cite{Wang2013}, might enable next generation of PIC with
nonlinearity at aJ-level ( 10 photons). Therefore, the QSE is promising
approach for future quantum state generation in PIC.

The idea of QSE can be extended from optical frequency to microwave
and terahertz frequencies. In particular, the superconducting quantum
circuits is readily to realize the QSE \cite{Kirchmair2013,Devoret2013}.
The mechanism of the QSE is very general, can also be generalized
to other Bosonic excitations, such as surface plasmon, exciton-polariton,
magnon and phonon. For example, quantum single phonon sources are
possible by nonlinear mechanical oscillators \cite{Villanueva2013}.

\emph{Conclusion.- }We have demonstrated that the simple NMZI can
filtrate single photon state from a weak coherent state. Using cascaded
NMZI, we can reliably extract arbitrary quantum state from a strong
coherent state. Since our scheme only requires Kerr nonlinearity,
linear phase shifter and beam splitter, it can be implemented in superconducting
circuits, coupled optomechanical systems, as well as photonic integrated
circuits.

C.L.Z. thanks Hailin Wang and Hong-Wei Li for fruitful discussion.
This work is supported by the ``Strategic Priority Research Program(B)''
of the Chinese Academy of Sciences (Grant No. XDB01030200), National
Basic Research Program of China (Grant Nos. 2011CB921200 and 2011CBA00200).
LJ acknowledges support from the DARPA Quiness program, the ARO, the
AFSOR MURI, the Alfred P Sloan Foundation, and the Packard Foundation.

\bibliographystyle{osa}

\cleardoublepage{}

\newpage{}

\newpage{}

\newpage{}

\newpage{}

\newpage{}

\newpage{}

\newpage{}

\onecolumngrid
\renewcommand{\thefigure}{S\arabic{figure}}
\setcounter{figure}{0}
\renewcommand{\thepage}{S\arabic{page}}
\setcounter{page}{1}
\renewcommand{\theequation}{S.\arabic{equation}}
\setcounter{equation}{0}
\setcounter{section}{0}

\begin{center}
\textbf{\textsc{\LARGE{}Supplemental Material for }}
\par\end{center}{\LARGE \par}

\begin{center}
{\Large{}``}\emph{\LARGE{}Filtration and Extraction of Quantum States
from Classical Inputs}{\Large{}''}
\par\end{center}{\Large \par}

\medskip{}

\begin{center}
Chang-Ling Zou$^{1,2,3}$, Liang Jiang$^{2}$, Xu-Bo Zou$^{1,3}$,
and Guang-Can Guo$^{1,3}$
\par\end{center}

\begin{center}
$^{1}$\emph{Key Lab of Quantum Information, University of Science
and Technology of China, Hefei 230026 }
\par\end{center}

\begin{center}
$^{2}$\emph{Department of Applied Physics, Yale University, New Haven,
CT 06511, USA }
\par\end{center}

\begin{center}
$^{3}$\emph{Synergetic Innovation Center of Quantum Information \&
Quantum Physics, University of Science and Technology of China, Hefei,
Anhui 230026, China}
\par\end{center}

\tableofcontents{}

\clearpage{}

\pagebreak{}

\section{Limitation of Two-Unit Quantum State Filtration}

\begin{figure}[H]
\centerline{ \includegraphics[width=0.9\textwidth]{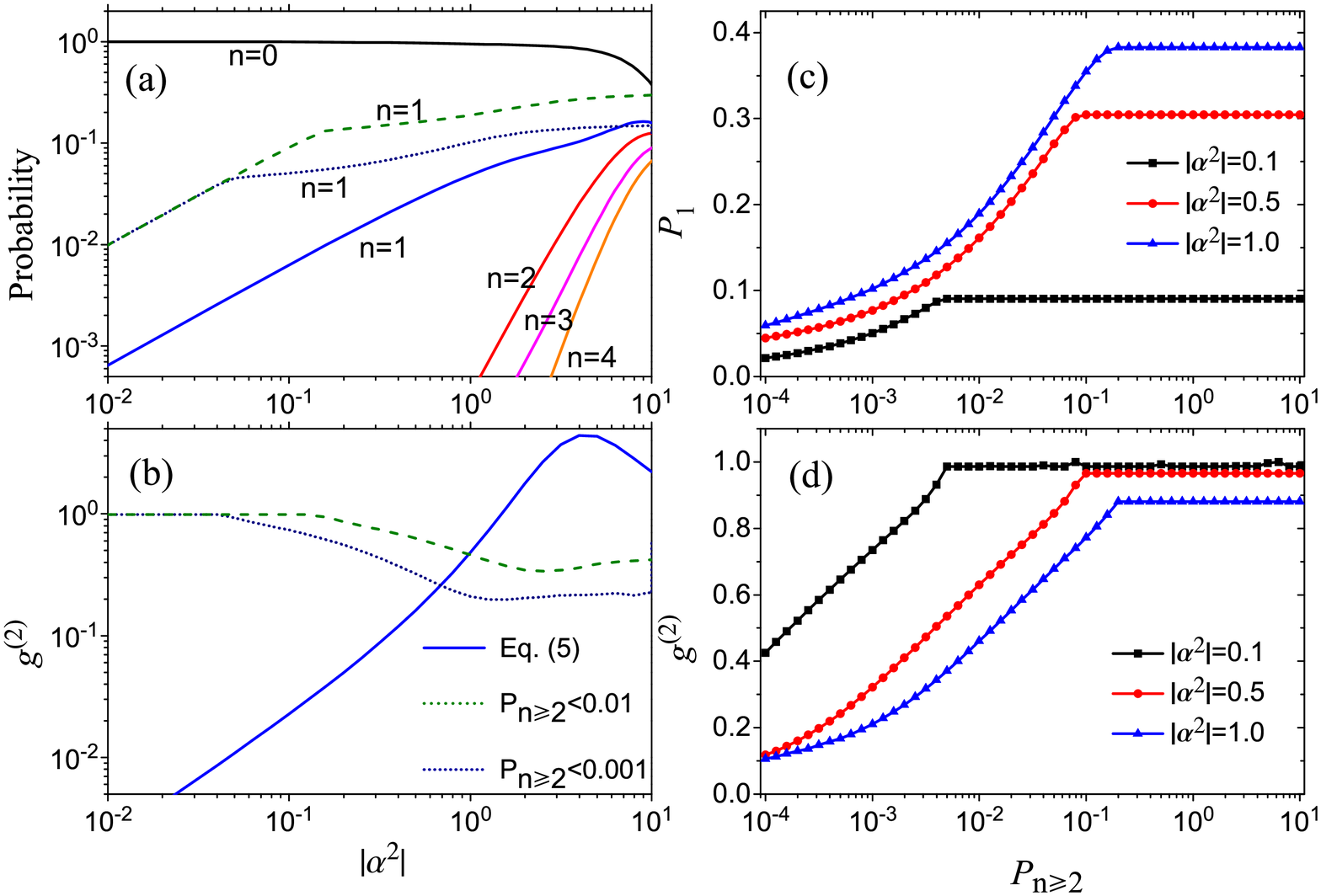}}\protect\caption{The probability of $n$-photon (a) and the photon statistic $g^{(2)}$
(b) at output for increasing input coherent laser intensity $|\alpha^{2}|$.
The solid lines are results for optimal conditions from Eq.\ (\ref{eq-optimal}),
the dashed lines are numerically optimized results for constrains
that $P_{n\geq2}<0.01$ and $P_{n\geq2}<0.001$. The dependence of
$P_{1}$ (c) and $g^{(2)}$ (d) against the optimization constrain
$P\geq2$. }

\label{LargeIntensity}
\end{figure}

Since the fidelity of single photon filtration (the brightness of
the single photon source) is limited as $P_{1}<\varphi|\alpha^{2}|/2$,
here we study the performance of the QSF with increasing input power
$|\alpha^{2}|$ numerically. Shown in Fig.$\,$\ref{LargeIntensity}(a)
by solid lines, the probabilities of Fock state outputs $P_{n}$ ($n=0,1,2,3,4$)
against $|\alpha^{2}|$ is calculated, for $\varphi=0.1$ and other
parameters satisfy the optimal condition Eq.\ (\ref{eq-optimal}).
For weak input, the $P_{1}$ increases linearly with $|\alpha|^{2}$
as expected, and $g^{(2)}$ also increases {[}Fig.$\,$\ref{LargeIntensity}(b){]}.
This means that the practical performance of the SPF deviated from
the expected perfect single photon source that $g^{(2)}\approx0$
for $|\alpha^{2}|\geq1$. The imperfection for larger input coherent
laser intensity should be attributed to the contributions of multiple
photons with $n\geq3$. This can be inferred from the probability
of $P_{2}$ and $P_{3}$ in Fig.$\,$\ref{LargeIntensity}(a), where
the $P_{2}$ is approximately linearly depends on $P_{3}$. This indicating
that the 2-photon output comes from the 3-photon component of input
state, which can't be eliminated efficiently by two-unit QSF.

Since the derivation of the optimal conditions is based on the assumption
that $|\alpha^{2}|\ll1$, the optimal condition may not be valid when
increase the input coherent laser power. Therefore, to gain $P_{1}$
as large as possible but keep the multiple photon probability as small
as possible, we may optimize the parameters of QSF numerically for
larger $|\alpha^{2}|$. For example, we optimize the $P_{1}$ under
the constrain $P_{n\geq2}<0.01(0.001)$, the results are shown by
dashed (doted) line in Figs. \ref{LargeIntensity}(a) and (b). For
$|\alpha^{2}|\geq1$, the performance of the QSF in both $P_{1}$
and $g^{(2)}$ are improved. For example, with the input mean photon
number $|\alpha|^{2}=1$, $P_{1}=0.047$ and $g^{(2)}=0.48$ for the
optimal condition from Eq.\ (\ref{eq-optimal}). With numerical optimization,
we obtained the improved performance as $P_{1}=0.18$ and $g^{(2)}=0.48$
for constrain $P_{n\geq2}\leq0.01$ and $P_{1}=0.10$ and $g^{(2)}=0.22$
for constrain $P_{n\geq2}\leq0.001$. It's obvious that, As illustrated
in Fig.\ 2(c) and (d) by comparing the optimized $P_{1}$ and $g^{(2)}$
for different constrain $P_{n\geq2}$, there is a trade-off relation
between $P_{1}$ and $g^{(2)}$ due to the lack of ability to suppress
2-photon output from multiphoton input state.

\section{The Dependence on Kerr nonlinearity}

\begin{figure}[H]
\centerline{ \includegraphics[width=0.9\textwidth]{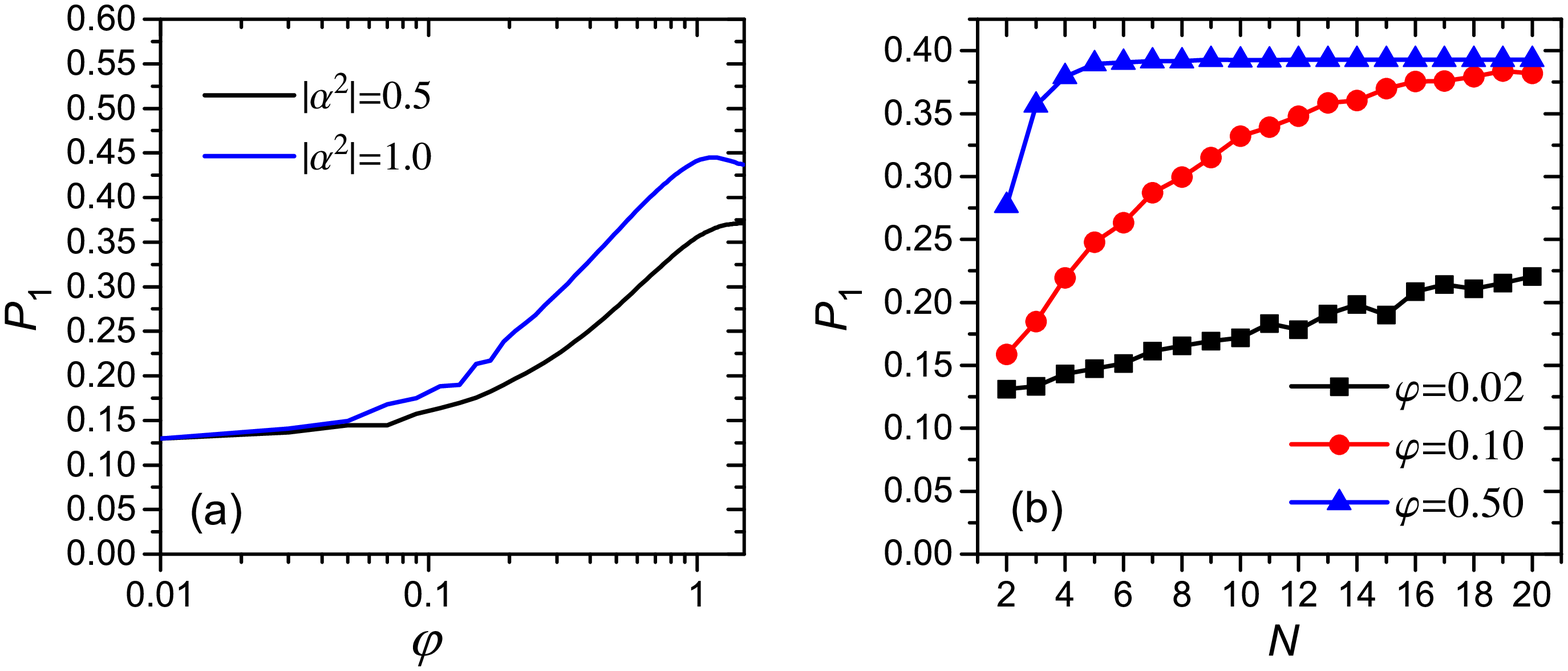}}\protect\caption{(a) The probability of single photon filtration by QSF against Kerr
nonlinearity $\varphi$ for different input coherent laser intensity
$|\alpha^{2}|$. (b) The probability of single photon extraction by
$N$-unit QSE against unit number $N$ for different $\varphi$. All
results are optimized under constrains that $P_{n\geq2}<0.01$.}

\label{Kerr}
\end{figure}

In the main body of the paper, most results of QSF are studied for
fixed nonlinearity parameter $\varphi=0.1$. From the results of QSF,
the fidelity of single photon filtration is also limited by the Kerr
nonlinearity in the NMZI. Therefore, we provide more results to study
the dependence of the performance of QSF/QSE on $\varphi$.

Shown in Fig. \ref{Kerr}(a) are the fidelity of single photon filtration
bu QSF against $\varphi$ for different $|\alpha^{2}|$. For both
$|\alpha^{2}|=0.5$ and $|\alpha^{2}|=1.0$, the $P_{1}$ increase
with $\varphi$, confirms that the asymptotic formula $P_{1}<\varphi|\alpha^{2}|/2$,
indicating that the performance of QSF by the simple NMZI can be improved
by larger nonlinearity. By further increase $\varphi$, $P_{1}$ reaches
the saturation value when $\varphi\approx1$.

In Fig. \ref{Kerr}(b), the $P_{1}$ by $N$-unit QSE against $N$
for various $\varphi$ are studied. It's not surprising that the performance
of QSE is also improved by increasing $N$, and reaches the saturation
value $1-\left|\left\langle 0|\alpha\right\rangle \right|^{2}$. Comparing
different curves, the higher the nonlinearity is, the less units required
to reach the maximum fidelity.
\begin{figure}[H]
\centerline{ \includegraphics[width=0.45\textwidth]{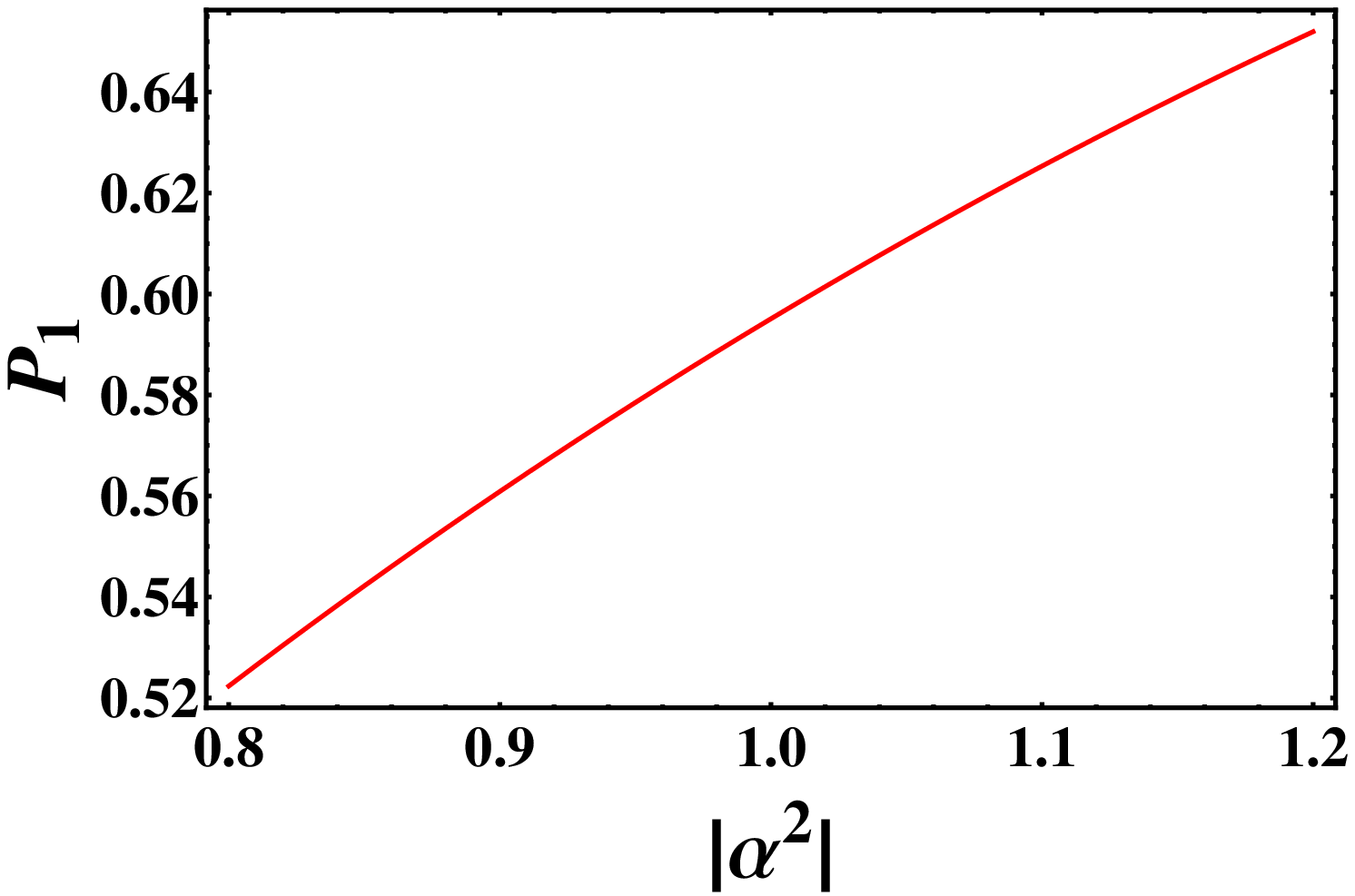}}\protect\caption{The $P_{2}$ for different input coherent laser power.}
\end{figure}

\section{Imperfections}

To test the performances of QSE against the imperfections of parameters,
we studied the performance of the single photon extraction with the
optimized parameters for the QSE consist of $N=20$ units with $|\alpha|^{2}=1.0$
and $\varphi=0.1$. In Fig.\ S3, the $P_{1}$ for varying input coherent
laser intensity is shown. In a wide range of input coherent state
mean photon number, the $P_{1}$ is not deviated much from the optimized
value.

In Fig.\ S4, the $P_{1}$ is tested with random perturbations of
$\phi$ and $\theta$ from the optimized value. The perturbations
are randomly and uniformly distributed in the range from $-0.01\pi$
to $0.01\pi$. From the results, the performance of the QSE show certain
tolerance to the parameter imperfections.

\begin{figure}[ptb]
\centerline{ \includegraphics[width=0.45\textwidth]{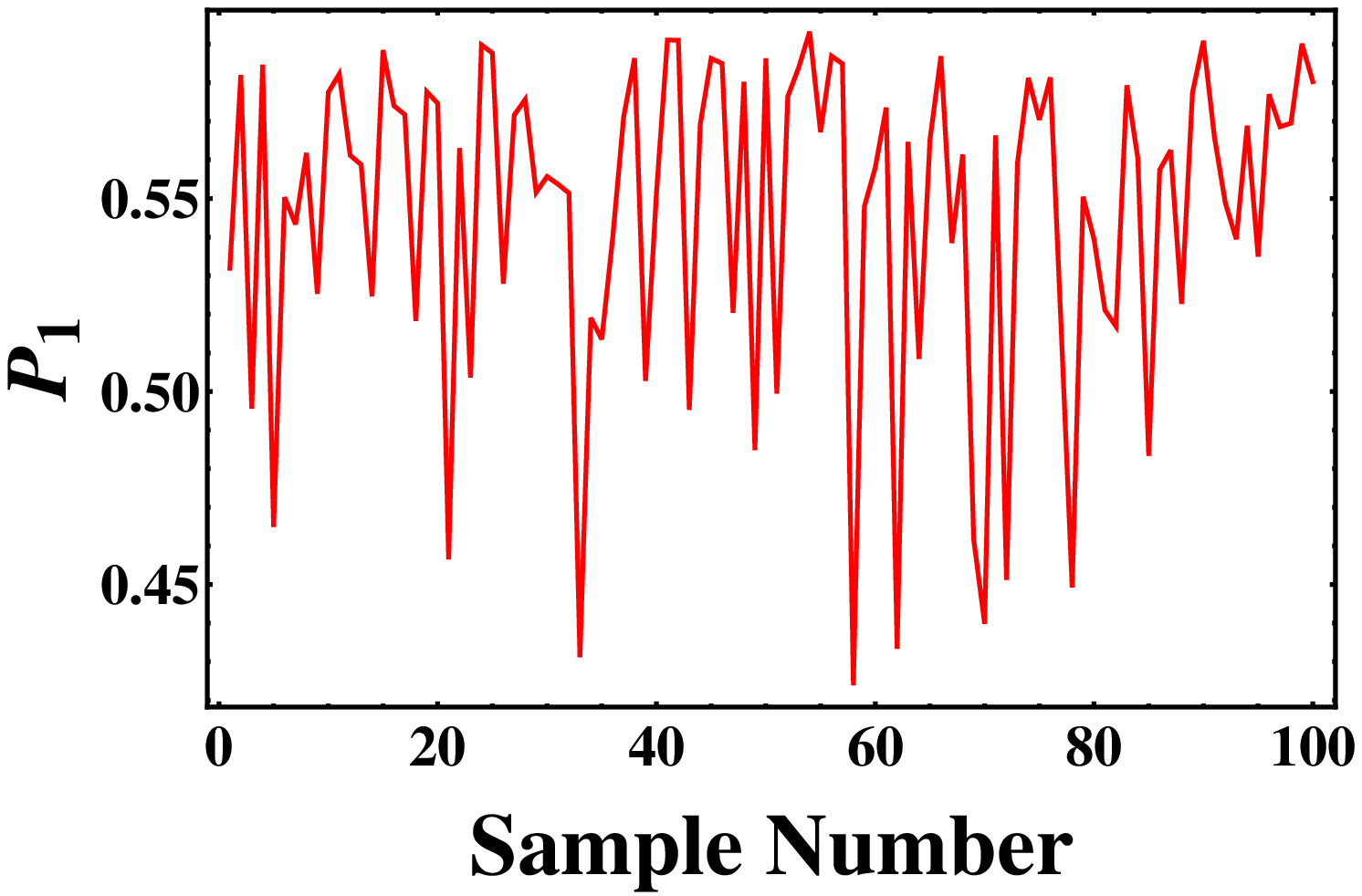}\includegraphics[width=0.45\textwidth]{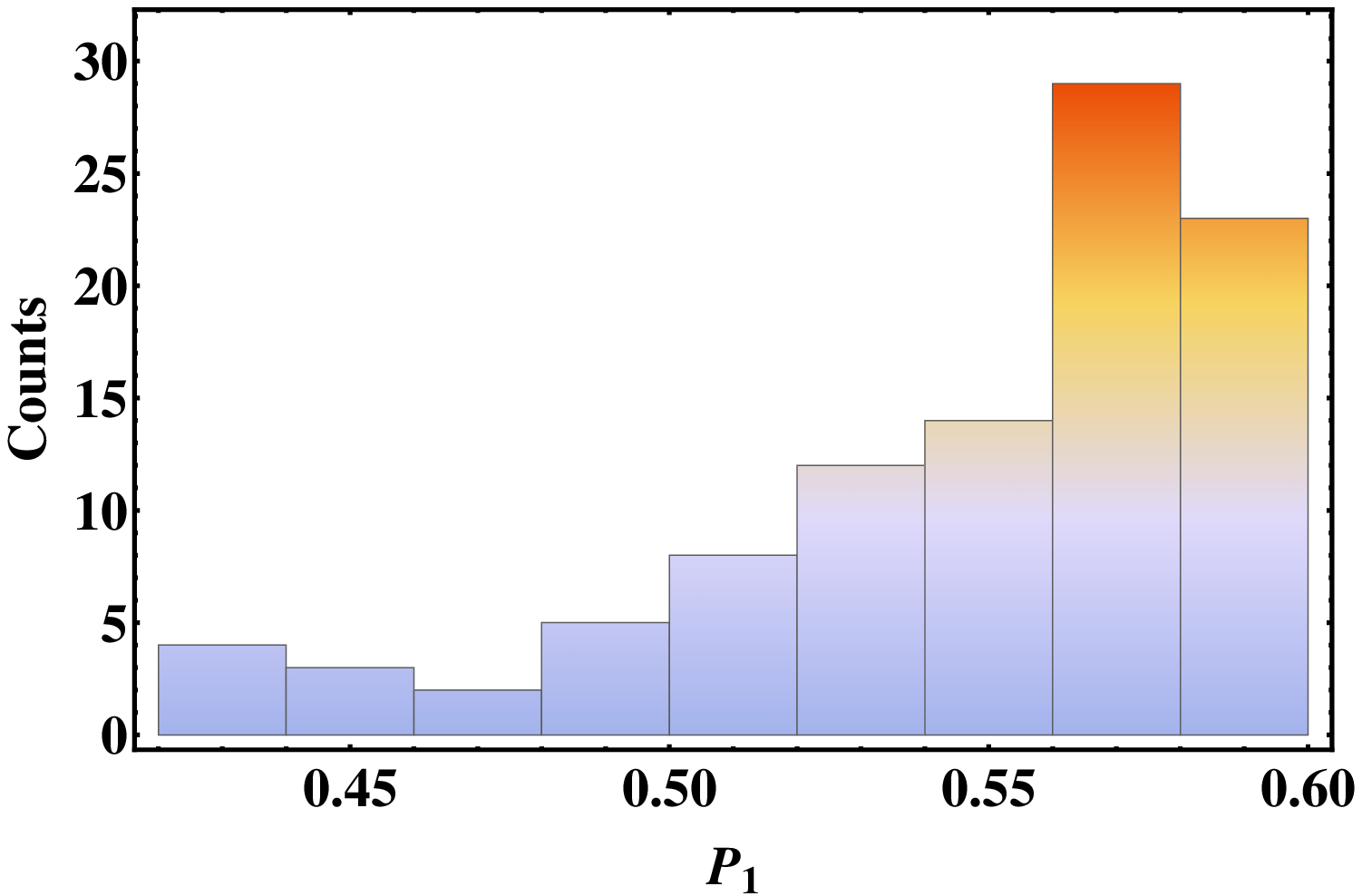}}\protect\caption{(Left) The $P_{1}$ for different samples of varied parameters. (Right)
The histogram of the $P_{1}$ for the random variation of parameters.}
\end{figure}

\section{Proof of High Fidelity}

As evidences shown in Figs.\ (2)\&(3), the purity and fidelity of
the QSF/QSE can approach unitary when $|\alpha|^{2}$ and $N$ approaches
infinity. Here, we provide the analytical analysis to support this
argument. In the example, the target state is
\begin{equation}
\left\vert \psi_{target}\right\rangle =(\left\vert 0\right\rangle +\left\vert 1\right\rangle )/\sqrt{2}.
\end{equation}
In general, the output state can be represent by
\begin{equation}
U\left\vert 0\right\rangle _{A}\left\vert \alpha\right\rangle _{B}=\left\vert 0\right\rangle _{A}\sum_{n=0}^{\infty}c_{n}\zeta_{0,n}\left\vert n\right\rangle _{B}+\left\vert 1\right\rangle _{A}\sum_{n=1}^{\infty}c_{n}\zeta_{1,n-1}\left\vert n-1\right\rangle _{B},
\end{equation}
where $U$ is unitary transformation, $c_{n}=\frac{\alpha^{n}}{\sqrt{n!}}e^{-\left\vert \alpha\right\vert ^{2}/2}$
for coherent input $\left\vert \alpha\right\rangle $, and the coefficients
satisfy$\left|\zeta_{0,n}^{2}\right|+\left|\zeta_{1,n}^{2}\right|=1$.
For $N\gg1$, it's possible to achieve near-perfect extraction of
Fock states that generate $\zeta_{0(1),n}=\frac{1}{\sqrt{2}}$ for
all $n$. Then, we obtain the fidelity of the QSE as
\begin{equation}
F=\frac{1}{2}(1+\sum_{n=0}^{\infty}c_{n+1}c_{n}^{*}).
\end{equation}
For $|\alpha^{2}|\gg1$, it can be approximated as
\begin{align}
F & \approx1-\frac{1}{16|\alpha|^{2}},
\end{align}
where
\begin{align}
\sum_{n=0}^{\infty}c_{n+1}c_{n}^{*}= & \sum_{n=0}^{\infty}\frac{\alpha}{\sqrt{n+1}}\frac{|\alpha|^{2n}}{n!}e^{-|\alpha|^{2}}\nonumber \\
\approx & \sum_{n=0}^{\infty}[1-\frac{n+1-|\alpha|^{2}}{2|\alpha|^{2}}+\frac{3}{8}\frac{(n+1-|\alpha|^{2})^{2}}{|\alpha|^{4}}]\frac{|\alpha|^{2n}}{n!}e^{-|\alpha|^{2}}\nonumber \\
= & 1-\frac{1}{8|\alpha|^{2}}+\frac{3}{8}\frac{1}{|\alpha|^{4}}.
\end{align}

\end{document}